\documentclass[a4paper]{VisionStyle} \usepackage{epsfig}

\begin{document}

\title{A fast X-ray timing capability on XEUS}

\author{D. Barret\inst{1} \and G. K. Skinner\inst{1} \and
E. Kendziorra\inst{2} \and R. Staubert\inst{2} \and L. Stella\inst{3}
\and M. Van der Klis\inst{4} }

\institute{ Centre d'Etude Spatiale des Rayonnements,
  CESR-CNRS/UPS,France \and University of Tubingen, Germany \and
  University of Roma, Italy \and University of Amsterdam, The
  Netherlands }

\maketitle

\begin{abstract} The Rossi X-ray Timing Explorer (RXTE) has
  demonstrated that fast X-ray timing can be used to probe strong
  gravity fields around collapsed objects and constrain the equation
  of state of dense matter in neutron stars. These studies require
  extremely good photon statistics. In view of the huge collecting
  area of its mirrors, XEUS could make a unique contribution to this
  field. For this reason, it has been proposed to include a fast X-ray
  timing capability in the focal plane of the XEUS mirrors. We outline
  the scientific motivation for such a capability, present some
  sensitivity estimates, and discuss briefly a possible detector
  implementation.
 \end{abstract}

\section{Introduction}

XEUS (X-ray Evolving Universe Spectroscopy) is proposed as a potential
follow-up mission to XMM-Newton. The primary task of XEUS will be to
perform spectroscopy of faint X-ray sources to trace the origins and
evolution of hot matter back to the early ages of the Universe (see
\cite{dbarret-G:hasingeretal00}). To achieve this goal, XEUS requires
a very large collecting area ($> 20$ m$^2$,
\cite{dbarret-G:bleeker00}).

The X-rays generated in the inner accretion flows aro\-und black holes
(BHs) and neutron stars (NSs) carry information about regions of the
strongly curved space-time in the vicinity of these objects. This is a
regime in which there are important predictions of general relativity
still to be tested. High resolution X-ray spectroscopy and fast timing
studies can both be used to diagnose the orbital motion of the
accreting matter in the immediate vicinity of the collapsed star,
where the effects of strong gravity become important. The
spectroscopic approach is already well covered by the XEUS detector
baseline, but the fast timing one should also be considered.  With the
discovery of millisecond aperiodic X-ray time variability (QPO) from
accreting BHs and NSs, and brightness burst oscillations in NSs, RXTE
has clearly demonstrated that fast X-ray timing has the potential to
measure accurately the motion of matter in strong gravity fields and
to constrain masses and radii of NSs, and hence the equation of state
of dense matter. With its huge collecting area, XEUS could provide an
order of magnitude sensitivity improvement in timing studies over
RXTE, if a fast X-ray timing capability is present in the focal plane.
In the following we outline the additional exciting science XEUS could
do with such a capability.

\section{Summary of scientific objectives}

\subsection{Probing strong gravity fields}
High-frequency QPOs have been seen in both BHs and NSs (see
\cite*{dbarret-G:vdk00} for an extensive review).  In BHs, the QPO
models proposed invoke General Relativity (GR) effects in the inner
accretion disk and depend strongly on the BH spin, making these QPOs
effective probes of spacetime near the event horizon (see e.g.
\cite{dbarret-G:m98}).  In NSs, three types of QPOs are commonly
observed, $\nu_{LF}$, 15--60\,Hz, $\nu_1$, 200--800\,Hz and $\nu_2$,
800-1200 Hz.  In the relativistic precession model
(\cite{dbarret-G:sv98}), ${\nu_{LF}, \nu_1, \nu_2}$ observed across a
wide range of objects are identified with three fundamental
frequencies characterizing the motion of matter in the strong field as
predicted by GR.  The low-frequency QPO at $\nu_{LF}$ is thought to be
due to {\it nodal precession}, dominated by the inertial-frame
dragging predicted by GR in the vicinity of a fast rotating collapsed
object.  The lower frequency kHz QPO at $\nu_1$ is identified with
relativistic {\it periastron precession}. Unlike the relativistic
effects seen in the orbit of Mercury and the relativistic binary
pulsar PSR1913+16, for which weak field approximations apply, the
periastron precession close to a collapsed star is dictated by strong
field effects. Finally $\nu_2$ is the orbital (``Keplerian'')
frequency; its value alone already restricts the allowed range of mass
and radius of the NS (\cite{dbarret-G:milleretal98}).  Other models
rely on a beat-frequency interpretation for both $\nu_1$ and
$\nu_{LF}$ (\cite{dbarret-G:as85,dbarret-G:milleretal98}), associate
the observed frequencies with oscillation modes of a precessing
accretion disk (e.g. \cite{dbarret-G:pn02}). Alternatively, in the
case of NSs, $\nu_1$ could be associated with a Keplerian frequency at
the outer radius of a boundary layer between the accretion disk and
the NS surface, $\nu_{LF}$ with radial oscillations in the boundary
layer, and $\nu_2$ with a frequency of a Keplerian oscillator under
the influence of the Coriolis force (\cite{dbarret-G:to99}).

With an order of magnitude better sensitivity, QPOs from a few Hz to
1600 Hz will be detected from many objects with a high enough
significance to use the data for crucial tests.  Regardless of the
physical origin of the QPOs at $\nu_{LF}$ and $\nu_1$, the increased
sensitivity and range will have dramatic benefits.  At higher
frequencies, either strong signatures of the innermost stable circular
orbit (ISCO), which sets an upper bound on $\nu_2$, will be discovered
from several sources (evidence has been found for one source so far:
4U1820-30, \cite{dbarret-G:zhangetal98}), or the frequencies
themselves will allow the elimination of several candidate equations
of state of dense matter (\cite{dbarret-G:milleretal98}).  In the {\it
  relativistic precession} interpretation, there are several
fundamental predictions yet to be tested.  First, the epicyclic
frequency $\Delta\nu=\nu_2-\nu_1$ should fall steeply to zero as
$\nu_2$ increases and the orbital radius approaches the ISCO. The
behaviour of the epicyclic frequency in the vicinity of BHs and NSs is
dominated by strong-field effects and drastically different from any
Newtonian or post-Newtonian approximation.  Hence it provides a
powerful test of the strong field properties of the metric (see
\cite{dbarret-G:sv99}).  According to the model $\Delta\nu$ should
also decrease for low values of $\nu_2$.  Second, $\nu_{LF}$ should
scale as $\nu_2^2$ over a wide range of frequencies (until
``classical'' terms due to stellar oblateness become important).
Observing such scaling would provide an unprecedented test of the
$1/r^3$ radial dependence of $\nu_{LF}$ predicted in the
Lense-Thirring interpretation (\cite{dbarret-G:sv99}).

In the above model, the QPO frequencies depend sensitively on the mass
(M) and angular momentum (J) of the compact star, as well as on the
radius at which the QPOs are produced.  M and J could be independently
estimated from waveform measurements at $\nu_{2}$, thus
overdetermining the problem so that the underlying theories can be
tested in critical ways.  The increased sensitivity of XEUS will
enable QPOs to be detected within their coherence times.  The cycle
waveform, which it will be possible to reconstruct, depends on the
Doppler shifts associated with the local velocity of the radiating
matter in the emitting blob or spot, as well as on curved-spacetime
light propagation effects.  If the frequency $\nu_2$ of the orbit is
known, QPO waveform fitting yields the mass $M$ (and Kerr spin
parameter) of the compact object.

\subsection{Equation of state of dense matter}
Nearly coherent oscillations at $\sim$ 300 Hz or $\sim$ 600 Hz have
been observed during type I X-ray bursts from about 10 NS so far (see
\cite{dbarret-G:s98} for a review).  These oscillations are probably
caused by rotational modulation of a hot spot on the stellar surface.
The emission from the hot spot is affected by gravitational light
deflection and Doppler shifts (e.g. \cite{dbarret-G:ml98}).  With
XEUS, the oscillation will be detected within one cycle. The
composition and properties of the NS cores have been the subject of
considerable speculation, and remain a major issue in modern physics :
at the highest densities, matter could be composed of pion or kaon
condensates, hyperons, quark matter, or strange matter.  By fitting
the waveform, it will be possible to investigate the spacetime around
the NS, and simultaneously constrain its mass and radius, and hence
determine the equation of state of its high density core (see e.g.
\cite{dbarret-G:nathetal02}).

\subsection{Additional science}
A fast X-ray timing capability would allow XEUS to investigate the
physics of a wide range of astrophysical sources, such as accreting
millisecond pulsars, micro-quasars, X-ray pulsars, dippers, CVs,
Novae, Soft gamma-ray re\-pea\-ters, Anomalous X-ray pulsars, \ldots.
For instance, there is only {\em one} accreting millisecond pulsar
known: SAXJ1808-3658 (\cite{dbarret-G:wv98,dbarret-G:cm98}).  Its
properties suggest that all NS systems should show pulsations at some
level.  In most models, pulse amplitudes cannot be suppressed below
$\sim0.1$\% (rms) without conflicting with spectroscopic or QPO
evidence.  With XEUS, the sensitivity to persistent millisecond
pulsations will be well below this level (pulsations at the level of
0.01\% rms would be detected in 1000 seconds in Sco X-1).  Detection
of such pulsations in objects also showing kHz QPOs and burst
oscillations would immediately confirm or reject several models for
these phenomena involving the NS spin (e.g.
\cite{dbarret-G:milleretal98}).  In addition, it has been suggested
that such objects could be among the brightest gravitational radiation
sources in the sky, emitting a periodic gravitational wave signal at
the star's spin frequency (\cite{dbarret-G:bildsten98}).  Undirected
searches in frequency space for such radiation lose sensitivity
because of statistical considerations.  Independently measuring the
spin period very accurately would therefore be of great importance for
periodicity searches with gravitational wave antennae (e.g.
\cite{dbarret-G:bradyetal97}).

For micro-quasars, the link between the very fast disk transitions
observed in X-rays and the acceleration process could be studied on
very short time scales, allowing the non steady state disk properties
and their link to the formation of relativistic jets to be explored
(\cite{dbarret-G:bellonietal97,dbarret-G:fenderetal99}).  This would
be of direct relevance to understanding the properties of AGNs, where
presumably similar jet formation mechanisms operate on a much larger
scales. In addition, through time-resolved spectroscopic observations,
the spacetime close to the black holes could be probed using the
variability of the iron K$\alpha$ line.

\section{XEUS sensitivity for timing studies}
\label{fauthor-E1_sec:tit} For the sensitivity computations, we have
assumed the energy response of the XEUS mirrors as given in the last
report of the telescope working group
(\cite{dbarret-G:aschenbachetal01}).  We have further assumed the
proposed high energy extension in which the inner mirror shells of the
telescope are coated with supermirrors (the effective area is thus
$\sim 20000$ cm$^2$ at $\sim 9$ keV and $\sim 1700$ cm$^2$ at 30 keV).
Finally we have assumed that the detector at the focal plane has a
detection efficiency equivalent to 2 millimeters of Silicon.  Table
\ref{dbarret-E1_tab:tab1} gives the count rates expected from some
sources.

\vspace*{-0.5cm}\begin{table}[h]  
\caption{Examples of total count rates above 0.5 keV and above 10
  keV (C$_{\rm E >10 keV}$) in kcts/s. The X-ray burst input spectrum is a
  blackbody of 1.5 keV with a normalization yielding an Eddington
  luminosity at 8.5 kpc. SAXJ1808-3659 is the millisecond pulsar taken at the peak of its 1996 outburst. }
  \label{dbarret-E1_tab:tab1}
  \begin{center}
    \leavevmode
    \footnotesize
    \begin{tabular}[h]{lccc}
      \hline \\ [-10pt]
      Source name & XEUS-1 & XEUS-2 & C$_{\rm E >10 keV}$ \\
      \hline \\ [-10pt]
      Crab & 253 & 811 & 5 \\
      Sco X-1 & 1210 & 3840 & 180 \\
      GC X-ray burst & 120 & 217 & 51 \\
     SAXJ1808-3659 & 34 & 135 & 2.5 \\
      \hline \\
      \end{tabular}
  \end{center}\vspace*{-1.0cm} \end{table}

Let us now compute the sensitivity to QPO and coherent signal
detections. First, for a QPO, the signal to noise ratio $n_\sigma$ at
which it is detected in a photon counting experiment is approximately:
$$n_\sigma = {1\over2}{S^2\over
  B+S}r_S^2\left(T\over\Delta\nu\right)^{1/2}$$
where $S$ and $B$ are
source and background count rate, respectively, $r_S$ is the (rms)
amplitude of the variability expressed as a fraction of $S$, $T$ the
integration time and $\Delta\nu$ the bandwidth of the variability. The
bandwidth $\Delta\nu$ is related to the coherence time $\tau$ of a QPO
as $\Delta\nu=1/\tau$. On the other hand, for a coherent signal
($T$$<$$1/\Delta\nu$), the more familiar exponential detection regime
applies, with false-alarm probability
$\sim$$\exp[-{S^2r_S^2T/2(B+S)}]$.

\begin{figure}[!t]
  \begin{center}
    \epsfig{file=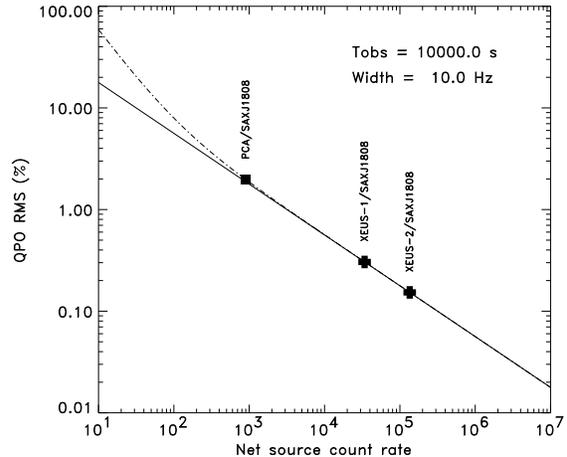,width=7.50cm}
  \end{center} 
\vspace*{-0.5cm}
\caption{Comparison between the XEUS (solid line) and
  RXTE/PCA (dot-dashed line) sensitivity for QPO detection ($5\sigma$
  in 10 ksec, signal width 10 Hz). An illustrative example is provided
  by the millisecond pulsar for which RXTE failed to detect QPOs. As
  can be seen, a factor of $\sim 10$ improvement in sensitivity over
  the RXTE/PCA is obtained with XEUS.}
\label{dbarret-G_fig:fig1}
\end{figure}

\begin{figure}[!ht]
\vspace*{-0.5cm}
  \begin{center}
    \epsfig{file=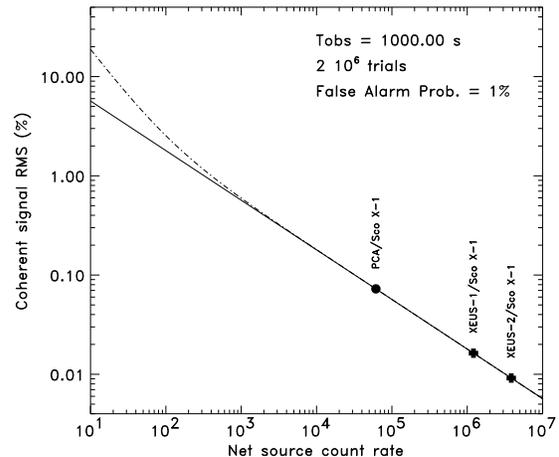,width=7.50cm}
  \end{center}
 \vspace*{-0.5cm}
 \caption{Comparison between the XEUS (solid line) and
   RXTE/PCA (dot-dashed line) sensitivities for coherent signal detection
   (1 ksec). The detection level corresponds to a false alarm
   probability of 1\% for $2\times 10^6$ trials. So far, no pulsations
   have ever been detected from Sco X-1.  The XEUS sensitivity is 10
   times better than the current RXTE/PCA sensitivity.}
  \label{dbarret-G_fig:fig2}
\end{figure}

From the above formulae, assuming B $\sim 0$ appropriate for XEUS, one
can estimate the RMS amplitude corresponding to a $5\sigma$ QPO
detection as a function of the source count rate (Figure
\ref{dbarret-G_fig:fig1}). Similarly one can compute the RMS for the
detection of a coherent signal at a given false alarm probability
(Figure \ref{dbarret-G_fig:fig2}). These two plots demonstrate that
with its huge collecting area XEUS provides an order of magnitude
sensitivity improvements in timing studies over RXTE.
\section{Detector implementation}
The detector needs to be able to handle up to 3 Mcts/s (XEUS-1) and 10
Mcts/s (XEUS-2) (equivalent to a 10 Crab source, see Table 1) with a
timing resolution of $\sim 10 \mu$s and a deadtime less than $\sim
0.1\mu$s.  Doing fast timing with good spectral resolution would
strongly increase the sensitivity for timing studies. The energy
resolution of the detector should therefore be good, ideally around
200 eV (i.e. a factor of $\sim 10$ improvement over the RXTE/PCA).
Finally the detector energy range should match closely the high energy
response of the mirrors.

\vspace*{-0.25cm}

In the current XEUS detector baseline, the Wide Field Imager (WFI) has
the highest count rate capabilities.  However, even in the most
optimistic case, it will only be able to provide timing information up
to 500 kcts/s (by using a fast window mode).  This means that an
alternative solution should be considered.  Among the fast X-ray
detectors currently available, Silicon Drift Detectors (SDDs) are the
most promising (\cite{dbarret-G:lechneretal01}).  The fast X-ray
capability for XEUS could be implemented with either a single SDD at
the focus, or a matrix of a few ($\sim 10$) SDDs placed out of focus.

\begin{figure}[!ht]
  \begin{center}
\epsfig{file=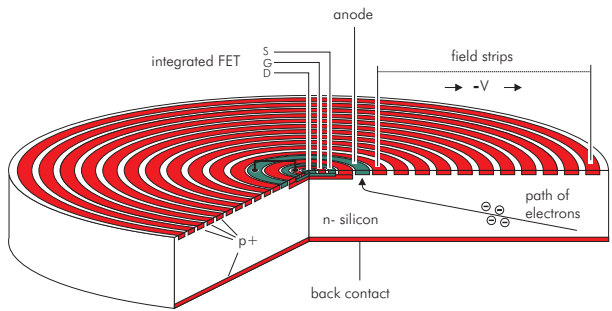,width=8.00cm}
  \end{center} \caption{Schematic cross section of a cylindrical
    Silicon Drift Detector (SDD). Electrons are guided by an electric
    field towards the small collecting anode located at the center of
    the device. The first transistor of the amplifying electronics is
    integrated on the detector ship (drawing kindly provided by P.
    Lechner).}  \label{dbarret-G_fig:fig3} \end{figure}

The SDD is a completely depleted volume of silicon in which an
arrangement of increasingly negative biased rings drive the electrons
generated by the impact of ionising radiation towards a small readout
node in the center of the device. The time needed for the electrons to
drift is much less than 1 $\mu$s.  The main advantage of SDDs over
conventional PIN diodes is the small physical size and consequently
the small capacitance of the anode, which translates to a capability
to handle high count rates simultaneously with good energy resolution.
To take full advantage of the small capacitance, the first transistor
of the amplifying electronics is integrated on the detector chip (see
Fig. \ref{dbarret-G_fig:fig3}). The stray capacitance of the
interconnection between the detector and amplifier is thus minimized,
and furthermore the system becomes practically insensitive to
mechanical vibrations and electronic pickup.

\vspace*{-.250cm}
\vspace*{-0.50cm}

With short signal shaping times (250 ns), SDDs have been proved to be
capable of handling count rates exceeding 1 Mcts/s with moderate
pile-up. Energy resolution of better than $\sim 200$ eV (at 6 keV,
equivalent to a low energy threshold $\sim 0.5$ keV) is readily
achieved with low cooling (-20\degr C) for count rates below 10$^5$
cts/s (e.g.  \cite{dbarret-G:lechneretal01}). With such a device, the
fast timing capability would explore completely new windows of X-ray
timing; first in the energy domain by getting below $\sim 2.5$ keV
(current threshold of RXTE-like proportional counters) and in the
frequency domain by reaching $\sim 10^4$ Hz (where signals have been
predicted, \cite{dbarret-G:sr00}).  SDDs are currently produced with
thicknesses of 300 microns. Although there are on-going efforts to
thicken these devices, the best match of the high energy response of
the mirror could be achieved by associating the SDD detector with a
higher density semi-conductor detector located underneath (e.g. CdTe,
CdZnTe, GaAs).  The device could be implemented independently of the
WFI or integrated on the sides of it as a separate detector.

The requirements in terms of telemetry rate, power, mass, volume and
cooling for the fast timing capability do not appear constraining for
XEUS.
\section{Conclusions} A fast X-ray timing capability on XEUS 
would nicely complement its primary science, at low cost. Probing
strong gravity fields, constraining the equation of state of dense
matter, and more generally studying the brightest sources with fast
X-ray timing would become possible with XEUS. There are no technical
issues: fast X-ray detectors capable of handling the expected count
rates already exist.

\begin{acknowledgements} We are grateful to the following colleagues
  who are supporting the proposal for a fast X-ray timing capability
  for XEUS: J.L. Atteia, T. Belloni, H. Bradt, L. Burderi, S. Campana,
  A. Castro-Tirado, D. Chakrabarty, P. Charles, S. Collin, S. Corbel,
  C. Done, G. Dubus, M. Gierlinski, J. Grindlay, A. Fabian, R. Fender,
  E. Gourgoulhon, J.M. Hameury, C. Hellier, W. Kluzniak, E. Kuulkers,
  S. Larsson, J.P. Lasota, D. de Martino, K. Menou, C. Miller, F.
  Mirabel, J.F. Olive, S. Paltani, R. Remillard, J. Rodriguez, R.
  Rothschild, T. di Salvo, M. Tagger, M. Tavani, L. Titarchuk, G.
  Vedrenne, N.  White, R. Wijnands, J. Wilms, A. Zdziarski, W. Zhang.
  
  The authors are grateful to A. Parmar for helpful information about
  the XEUS mission in general. We also thank G. Hasinger, L.  Strueder
  and P.  Lechner for discussions about the wide field imager and
  silicon drift detectors.  \end{acknowledgements}

\end{document}